\documentclass{emulateapj}

\usepackage{color}
\shorttitle{High brightness temperature in 3C~120}
\shortauthors{Roca-Sogorb et al.}

\begin{document}

\title{Unexpected high brightness temperature 140 pc from the core in the jet of 3C~120}

\author{Mar Roca-Sogorb\altaffilmark{1}, Jos\'e L. G\'omez\altaffilmark{1}, Iv\'an Agudo\altaffilmark{1}, Alan P. Marscher\altaffilmark{2} and Svetlana G. Jorstad\altaffilmark{2}}

\altaffiltext{1}{Instituto de Astrof\'{\i}sica de Andaluc\'{\i}a, CSIC, Apartado 3004, 18080 Granada, Spain. mroca@iaa.es; jlgomez@iaa.es; iagudo@iaa.es}
\altaffiltext{2}{Institute for Astrophysical Research, Boston University, 725 Commonwealth Avenue, Boston, MA 02215, USA. marscher@bu.edu; jorstad@bu.edu}

\begin{abstract}
  We present 1.7, 5, 15, 22 and 43 GHz polarimetric multi--epoch VLBA observations of the radio galaxy 3C~120. The higher frequency observations reveal a new component, not visible before April 2007, located 80 mas from the core (which corresponds to a deprojected distance of 140 pc), with a brightness temperature about 600 times higher than expected at such distances. This component (hereafter C80) is observed to remain stationary and to undergo small changes in its brightness temperature during more than two years of observations. A helical shocked jet model -- and perhaps some flow acceleration --  may explain the unusually high T$_b$ of C80, but it seems unlikely that this corresponds to the usual shock that emerges from the core and travels downstream to the location of C80. It appears that some other intrinsic process in the jet, capable of providing a local burst in particle and/or magnetic field energy, may be responsible for the enhanced brightness temperature observed in C80, its sudden appearance in April 2007, and apparent stationarity.
\end{abstract}


\section{Introduction}
  3C~120 is an active and relatively nearby (z=0.033) radio galaxy with a blazar--like one--sided superluminal radio jet that has proven to be an excellent laboratory for studying the physics of relativistic jets in active galactic nuclei \citep[e.g.,][]{Walker87,Walker01,Gomez98,Gomez99,Gomez00,Gomez01,Gomez08,Homan01,Marscher02,Marscher07,Jorstad05,Jorstad07, Chatterjee09, Marshall09}. Previous observations using the Very Long Baseline Array (VLBA) at high frequencies (15, 22 and 43 GHz) have revealed a very rich inner jet structure, containing multiple superluminal components as well as evidence for stationary features suggestive of a helical pattern viewed in projection \citep{Walker01,Hardee05}.

\section{Observations and data reduction}
 
\begin{figure*}
\centering
\includegraphics[scale=0.80]{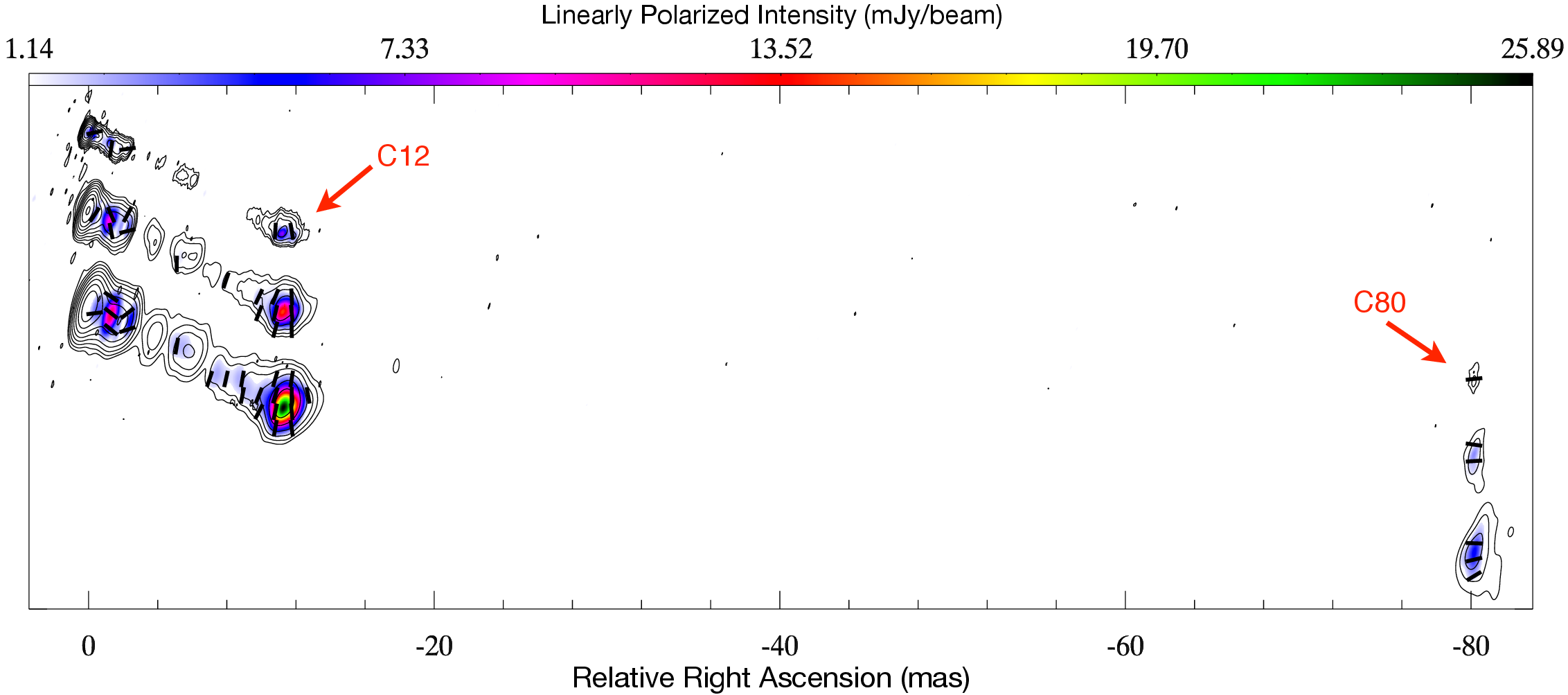}
\caption{VLBA images of 3C~120 in 2007 November 7 (2007.85) at 43 ({\it top}), 22 ({\it middle}) and 15 ({\it bottom}) GHz. Ten logarithmic contours are plotted for the total intensity images between the first contour at 0.18, 0.16, and 0.08\%, and the last contour at 90\% of the peak brightness of 1.11, 1.27, and 1.12 Jy beam$^{-1}$ at 43, 22 and 15 GHz, respectively. Color shows the linearly polarized intensity, and bars indicate the direction of the electric polarization vector. A convolving beam of 0.63$\times$0.22, 1.25$\times$0.40, 1.76$\times$0.61 mas at -17.9, -18.5, and -17.9$^{\circ}$, was used at 43, 22 and 15 GHz, respectively.}
\label{3freq}
\end{figure*}

  We present VLBA observations taken in November 2007, as part of a multi--frequency program to map the rotation measure in 3C~120 at all accessible VLBA scales, and in February 2001, when the VLBA was used as part of the ground array for HALCA observations of 3C~120 at 5 GHz.

  The November 2007 VLBA observations were performed at 43, 22, 15, [4.6--5.1] and [1.35--1.75] GHz in dual polarization, with 9 antennas of the VLBA (Saint Croix was down for maintenance). The highest frequency observations were performed with 32 MHz continuous bandwidth centered at the standard 43, 22 and 15 GHz frequencies. The 4 and 20 cm receivers were split into four 8 MHz bandwidths to maximize possible detection of low Faraday rotation measure.
 
  Reduction of the data was performed with the AIPS software in the usual manner \citep[e.g.,][]{Leppanen95}. The absolute phase offset between the right-- and left--circularly polarized data, which determines the electric vector position angle (EVPA), was obtained by comparison of the integrated polarization of the VLBA images of several calibrators (0420$-$014, DA193, 3C~279, 3C~454.3 and 4C~39.25) with simultaneous VLA observations, as well as archival data from the UMRAO, MOJAVE and NRAO long term monitoring programs. Estimated errors in the orientation of the EVPAs lie in the range of 5$^{\circ}$-10$^{\circ}$. After the initial reduction, the data were edited, self--calibrated, and imaged both in total and polarized intensity with a combination of AIPS and DIFMAP \citep{Pearson94}.

\section{High brightness temperature in 3C~120}

\begin{figure}
\centering
\includegraphics[scale=0.50]{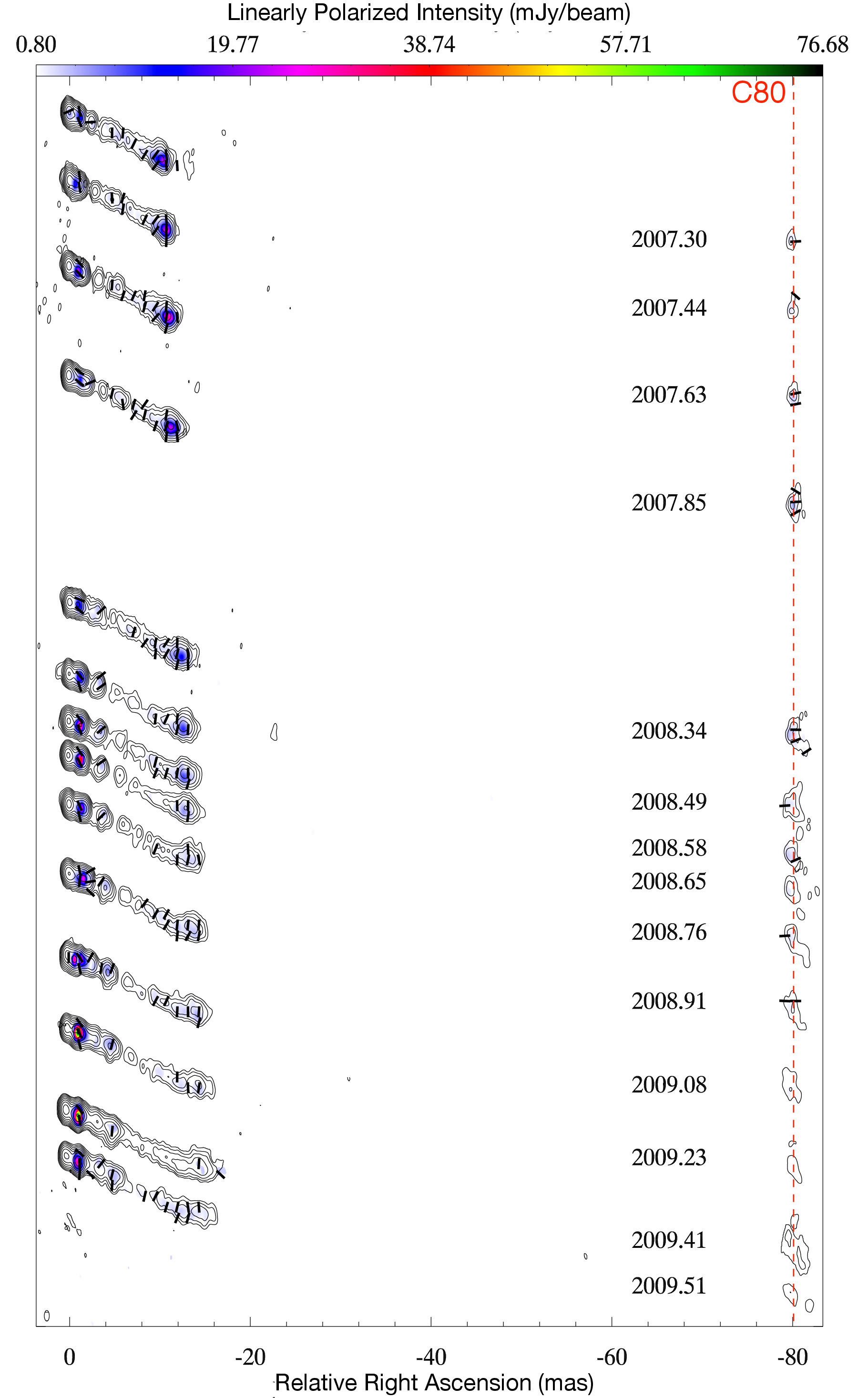}
\caption{Multi--epoch VLBA images of 3C~120 at 15 GHz from our 2007.85 observations and the MOJAVE program. Vertical map separation is proportional to the time difference between successive epochs of observation, shown for each image. Total intensity is plotted in contours at 1, 2, 4, 8, 16, 32, 64, 128, 256, 512, and 1024 mJy beam$^{-1}$. Polarized intensity is shown in color scale and bars (of unit length) indicate the electric vector position angle. A common convolving beam of 1.18$\times$0.55 mas at 0$^{\circ}$ was used for all images, and is shown in the lower left corner. All of the images have been registered through comparison of the positions of several components.}
\label{Mojave}
\end{figure}

  Our high frequency (15--43 GHz) VLBA observations during November 2007 (Fig.~\ref{3freq}) reveal a component (hereafter C80) located 80 mas from the core \citep[deprojected to $>\!140$ pc for a viewing angle $<\!20^{\circ}$;][]{Gomez00}. This is an unusually large distance for detecting emission at these high frequencies --in fact, none of the previous VLBI observations of 3C~120 (starting in 1982) have ever reported emission at this distance at 5 GHz or higher frequencies. We have remapped our previous 15, 22 and 43 GHz VLBA data taken from 1996 to 2001\citep{Gomez98, Gomez99, Gomez00, Gomez01, Gomez08}, covering more than 30 epochs, to check whether we missed it in our previous analysis, but find no indication for emission at the region of C80. However, remapping of the 15 GHz data published in the MOJAVE database (containing also data from other programs; see \texttt{http://www.physics.purdue.edu/MOJAVE} for more information) revealed the first detection of C80 in April 2007. After this epoch C80 appears in all 15 GHz images, as shown in Fig.~\ref{Mojave}. This sequence of images shows no significant motions for C80 during the nearly two years covered by the observations. However, during 2007 the component is observed to increase in flux density and to remain quite compact. Later on, C80 becomes more extended, elongating in the south--west direction without significant changes in its flux. 

\begin{figure}
\centering
\includegraphics[scale=0.41]{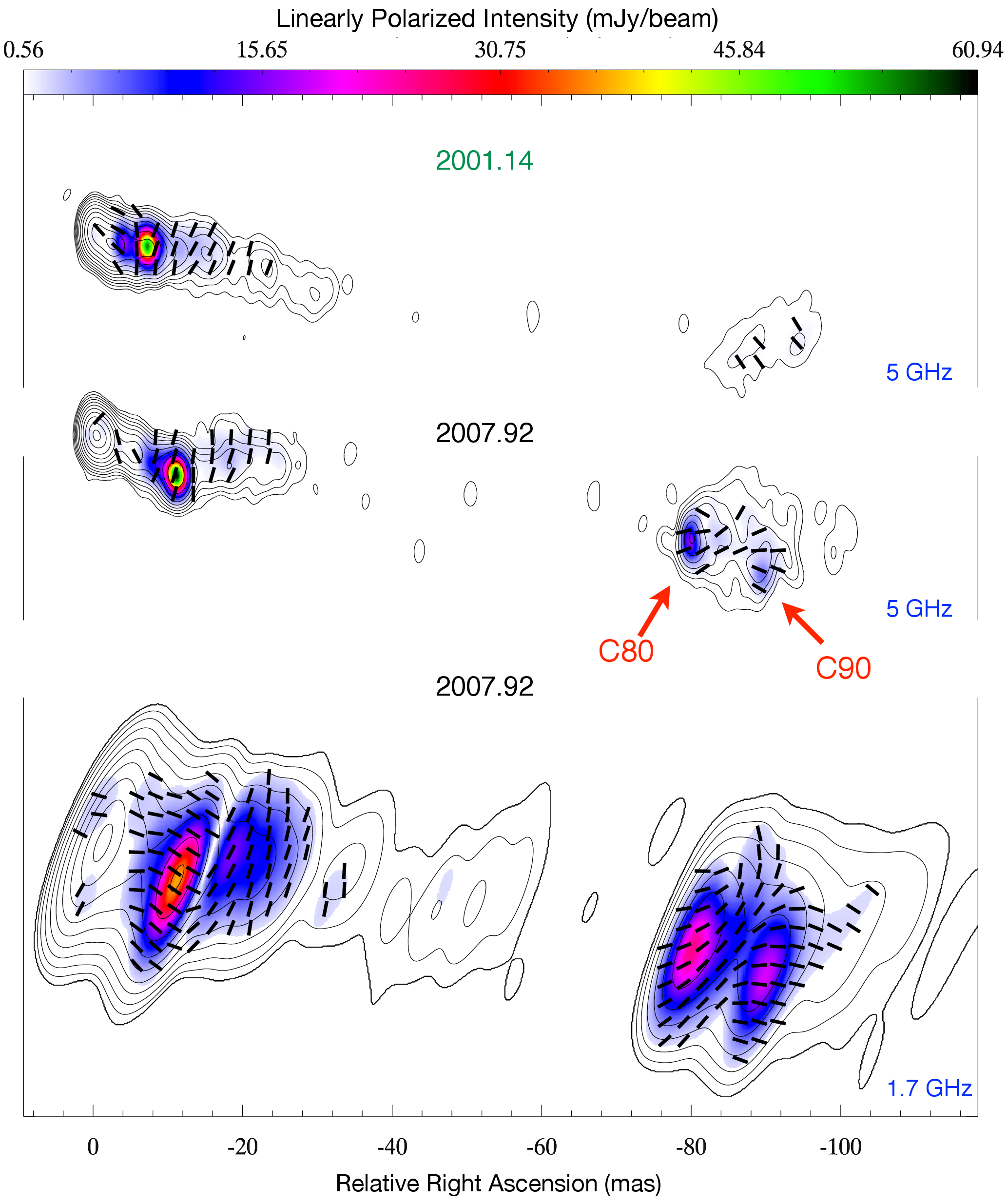}
\caption{VLBA observations of 3C~120 at 5 GHz in 2001 February 21 ({\it top}), and 2007 November 30 at 5 GHz ({\it middle}), and 1.7 GHz ({\it bottom}). Total intensity is plotted in contours at 0.1, 0.24, 0.48, 0.9, 1.8, 3.6, 7.2, 14.4, 28.8, 57.6 and 90\% of the peak brightness of 0.95 ({\it top}), 0.97 ({\it middle}), 0.84 ({\it bottom}) Jy beam$^{-1}$. Polarized intensity is shown in color scale and bars (of unit length) indicate the electric vector position angle. A common convolving beam of 3.58$\times$1.71 mas at 0$^{\circ}$ was used for the images at 5 GHz, and 13.70$\times$3.74 mas at -22.10$^{\circ}$ at 1.7 GHz.}
\label{VSOPvsBG182}
\end{figure}

\begin{figure}
\centering
\includegraphics[scale=0.20]{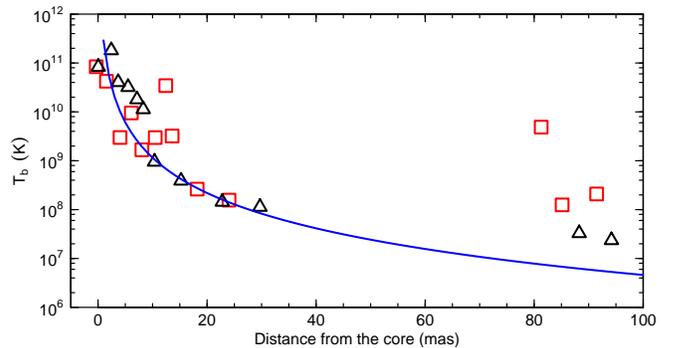}
\caption{Observed brightness temperatures for different components along the jet in the 5 GHz images in February 2001 (black) and November 2007 (red). The blue line represents the expected decline with distance from the core.}
\label{VSOPvsBG182Tb}
\end{figure}

  Our lower frequency images at 1.7 and 5 GHz (see Fig.~\ref{VSOPvsBG182}) show that the region located around C80 corresponds to a double structure, with another component at $\sim$90 mas (hereafter C90) located at the southernmost side of the jet, after which the jet extends to the northwest direction. This contrasts to what is found in our previous 2001 image at 5 GHz (see Fig.~\ref{VSOPvsBG182}), which shows extended emission located at $\sim$85--95 mas and elongated in the north--west direction. We have used circular Gaussian brightness distributions to model--fit the different components in the jet for both epochs, from which we have estimated their observed (i.e., uncorrected by Doppler boosting) brightness temperatures, T$_b$ (see Fig.~\ref{VSOPvsBG182Tb}). The T$_b$ along the jet is observed to decline with distance from the core following the $r^{-2.4}$ proportionality found by \cite{Walker87}. Component C80 has a brightness temperature of $5\times10^9$ K, which is about 600 times larger than the expected value of $\sim\!8\times10^6$ K at such large distance from the core. $\!8\times10^6$ K is also the typical detection threshold for VLBA observations at 5 GHz. Hence, the fact that C80 has not been detected in any of the previous 5 GHz VLBA images implies an increase in its brightness temperature by at least a factor of 600. This unusually high T$_b$ explains why C80 has become visible even at the highest VLBA observing frequencies (see Fig.~\ref{3freq}).

  Figures \ref{3freq}, \ref{Mojave} and \ref{VSOPvsBG182} show the electric vector position angle (EVPA) to be aligned with the local direction of the jet in C80 for observations after 2007, in contrast to what is found for the remainder of the jet \citep[see also][]{Walker01}, and to that shown in the 5 GHz image taken in 2001 (see Fig.~\ref{VSOPvsBG182}). Maps of the rotation measure at different frequency intervals during November 2007 (G\'omez et al., in preparation) show values of the order of 10 rad m$^{-2}$ for C80, small enough to marginally rotate only the EVPAs at 1.7 GHz. Hence, we can conclude that the observed magnetic field in C80 is perpendicular to the local intensity structure for observations after 2007. The degree of polarization of C80 is $\sim$20\% and the spectral index of the region is $\alpha\sim-1$ (S$_\nu$ $\propto$ $\nu^{\alpha}$), which is similar to the values found for the rest of the optically thin jet.

\section{Discussion}

  3C~120 has been extensively observed at 1.7 GHz by \citet{Walker01}, showing a variety of moving knots and a side--to--side structure suggestive of a helical pattern seen in projection \citep[see also][]{Hardee05}, in which the helical twisted flow along the southern side of the jet is more closely aligned with the line of sight. \cite{Walker01} identified a component located at 81 mas from the core that appeared to be stationary (between 1982 and 1997), and could correspond to one of the southernmost components produced by the enhanced differential Doppler boosting. We are therefore tempted to identify this with component C80. However, our low frequency observations show that at the location of the C80 component the emission structure changed significantly between 2001 and 2007, and that the southernmost emission in 2007 corresponds to C90, instead of C80 (see Fig.~\ref{VSOPvsBG182}). Hence, C90 would be associated with a jet region flowing at a smaller viewing angle, and it is therefore very unlikely that C80 would correspond to another bend in the jet, given the estimated helical wavelengths \citep{Hardee05}. Furthermore, a bend in the jet would lead to an increase in T$_b$ by a factor of ($\delta_{new}/\delta_{old})^{n}$, where $\delta$ is the Doppler factor and $n$ is $2\!-\!\alpha$ for the case of continuous jet or $3\!-\!\alpha$ for a moving inhomogeneity \citep{Readhead94}. In our case of a bend in the jet and an estimated spectral index of $\alpha=-1$ we have $n=3$; therefore to account for the 600 increase in T$_b$ it is required an increase in $\delta$ by a factor of $\sim$8.4 with respect to the estimated mean Doppler factor $\delta_{old}=2.4$ \citep[corresponding to a Lorentz factor $\gamma_{old}=5.3$ and viewing angle $\theta_{old}=20^{\circ}$; ][]{Jorstad05}. This involves an unlikely acceleration of the jet from a Lorentz factor of 5.3 to $\sim$10.1, even for the most favorable case of a jet pointing directly towards the observer.

\begin{figure}
\includegraphics[scale=0.5]{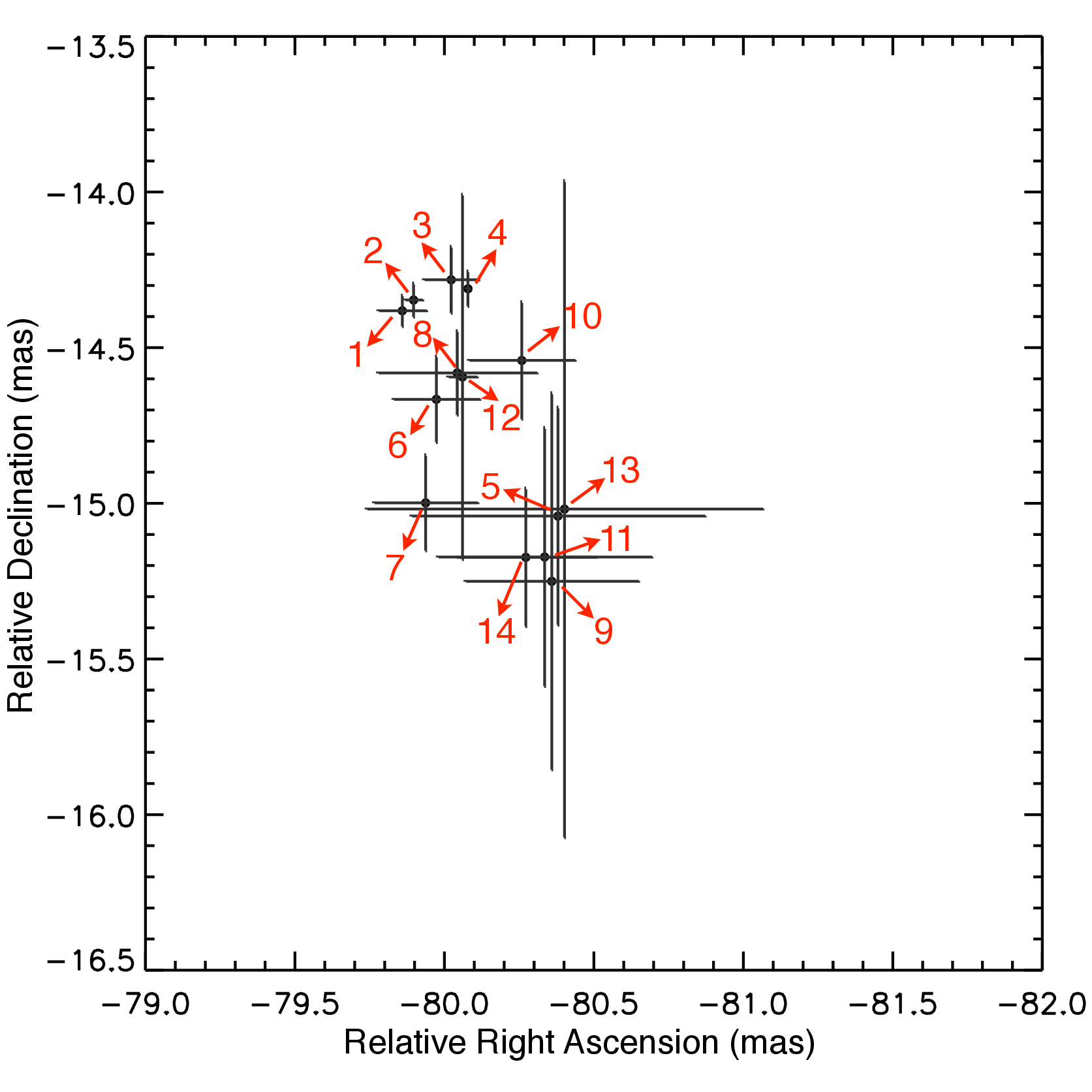}
\caption{Position of component C80 at different epochs. An elliptical Gaussian brightness distribution has been used to model--fit the C80 component in the VLBA images at 15 GHz from 2007.30 to 2009.51 (see Fig.~\ref{Mojave}). Labels (from 1 to 14) indicate the epochs of observation in chronological order. The errors in the position have been calculated as the standard deviation of the position obtained by using task \textsf{MFIT} in AIPS and model--fitting with DIFMAP.}
\label{MojavePosition}
\end{figure}

  Could C80 instead correspond to a moving shock whose motion through a bend towards the observer has resulted in an apparently stationary feature? The effect of a shock is determined by the compression factor, $\eta$, so that the magnetic field is scaled up as $B\rightarrow B/\eta$ and the electron energy density as $N_{0} \rightarrow N_{0} \eta^{-(\gamma+2)/3}$, where $\gamma$ is the electron energy spectral index \citep[e.g.,][]{Hughes89,Gomez93}. This yields an increase in the optically thin specific intensity of synchrotron radiation by a factor of $\eta^{-(5\gamma+7)/6}$, or equivalently $\eta^{(5\alpha-6)/3}$. The most likely scenario would then involve a shocked helical jet, in which the effects of the shock wave compression and the differential Doppler boosting would add to produce an increase in the brightness temperature by a factor of ($\delta_{new}/\delta_{old})^{3-\alpha} \eta^{(5\alpha-6)/3}$. Note that for a moving shock we use $n=3\!-\!\alpha$, on the assumption that the radiating fluid is moving close enough to the shock speed. It is possible to obtain an upper limit to $\eta$ by maximizing the contribution from the Doppler boosting considering that in C80 the jet points directly towards the observer, but maintains the same Lorentz factor of 5.3. The factor of 600 increase in T$_b$ for C80 would then require a relatively weak shock with $\eta\leq0.87$. This is in fact a too conservative value, since as mentioned previously C90 is at a smaller viewing angle than is C80, so that the jet cannot point directly towards the observer at C80. If the jet instead bends to a viewing angle of $5^{\circ}$ ($10^{\circ}$) then $\eta=0.71$ ($\eta=0.45$). For comparison, the component located at $\sim$12 mas (C12; see Fig.~\ref{3freq}) -- which is one of the most intense ever observed in 3C~120 -- has $\eta\!\sim\!0.35$. The unusually high T$_b$ of C80 could therefore be explained by a combination of jet bending and a moving shock --and perhaps also some jet flow acceleration and/or unusually large particle acceleration-- but it seems very unlikely that it corresponds to the usual shock that appears near the core and moves downstream to the location of C80: as simulated by \citet{Gomez94-2}, a component moving through a helical jet would progressively increase in flux as it approaches the bend, accompanied by a rotation of its EVPA. An increase in the flux density of C80 is indeed observed during 2007, but not later. Some motion of C80 would also be expected as it approaches the most favorably oriented jet region corresponding to C90, which contrasts with the quasi-stationarity of C80 shown in Fig.~\ref{MojavePosition}. Therefore, a shock moving through a helical jet cannot account entirely for the observed properties of C80.

  It appears that a strong, stationary shock generated in situ, at the location of C80, is needed. This can be a standing shock, produced perhaps by a steep decrease in the external pressure. As has been proposed to explain the flaring HST-1 knot in the M87 jet by \cite{Stawarz06}, the brightening of C80 in April 2007 may mark the arrival of excess particles and photons produced by the active nucleus in the past. For a jet flow Lorentz factor similar to that measured for the components, we can estimate that the core flare should have taken place near 1975. Indeed, the light curve of 3C~120 shows a period of very high activity at this epoch, corresponding to the maximum observed centimeter--wave flux since monitoring began in the mid--1960s (see \texttt{http://www.astro.lsa.umich.edu/}). It seems, however, difficult to explain why such a large injection of energy into the jet did not result in a shock that remained bright as it moved down the jet, visible as an intense moving component in any of the following VLBI observations.
  
  If component C80 corresponds to a fixed planar shock it is possible to estimate the minimum Lorentz factor change required to explain the observed brightness temperature change:
\[
  \frac{T_{b,new}}{T_{b,old}} = \left( \frac{\delta_{new}}{\delta_{old}} \right)^{2-\alpha}
                                                 \left( \frac{\eta_{new}}{\eta_{old}} \right)^{\frac{5\alpha-6}{3}}
                                                 > 600
\]
where $\eta=\gamma( 8 \gamma^4 - 17 \gamma^2 + 9 )^{-1/2} \gtrsim (\sqrt 8 \gamma)^{-1}$, and $\gamma$ is the upstream flow velocity \citep{Hughes89}. If we simplify by assuming that the jet points directly towards the observer we can write
\[
  \left( \frac{\delta_{new}}{\delta_{old}} \right)^{2-\alpha}
  \left( \frac{\eta_{new}}{\eta_{old}} \right)^{\frac{5\alpha-6}{3}}
                               \sim   
  \left( \frac{\gamma_{new}}{\gamma_{old}} \right)^{2-\alpha}
  \left( \frac{\gamma_{new}}{\gamma_{old}} \right)^{\frac{6-5\alpha}{3}}
\]
and thus
\[
\frac{T_{b,new}}{T_{b,old}} \approx \left( \frac{\gamma_{new}}{\gamma_{old}} \right)^{\frac{20}{3}} > 600
\]
which requires $\gamma_{new} > 2.6 \gamma_{old}$, that is, a flow acceleration from the estimated value of 5.3 to $>$13.8. Such an acceleration of the upstream flow is difficult to measure since VLBI observations provide an estimation of components pattern velocities, rather than actual bulk flow velocity, but some acceleration in the components may be expected in case of an accelerating jet \citep[e.g.,][]{Agudo01,Homan09}. However, a clear systematic acceleration of the components in 3C~120 has not been previously observed \citep{Walker87,Walker01,Gomez01,Homan01,Homan09,Jorstad05}.
  
  It is also possible that a moving component might be passing through a standing shock located at the position of C80, leading to the initial increase in flux. After the interaction, the two components would split, with the C80 stationary component associated with the standing shock progressively recovering its initial position, flux, and orientation of electric vector \citep{Gomez97}, while the moving component increases its flux as it approaches the bent region at C90.

  Component C80 could also result from a strong interaction with the external medium, similar to that proposed for the inner jet regions \citep{Gomez00}. However, the low values of the rotation measure (RM $\sim$10 rad m$^{-2}$) found for the C80--C90 region suggest that such interaction with the external medium is probably not taking place. 

  The observations made by \citet{Walker01}, covering 1982 to 1997, and those presented in this work (2001 and 2007), show evidence that, although the region located at $\sim$ 80--90 mas changes with time, it appears to correspond to a common bent region, where the jet is oriented more towards the observer. In this case the region located at 81 mas in 1997, identified by \cite{Walker01} as a stationary component, could in fact correspond to the southernmost region located at $\sim$ 86 mas in 2001, and this in turn to the C90 component seen in 2007. This motion of the bent jet region can be explained in the framework of a slowly moving helical pattern, as simulated by \citet{Hardee05}. The estimated upper limit of $\sim$ 0.55 mas yr$^{-1}$ \citep[$\sim$ 1.1 $c$;][]{Walker01} for the pattern speed of the helix is consistent with the observations between 2001 and 2009.
  
  Although the helical jet model can explain the observed properties of C90, none of the proposed models provides a complete explanation for the unusually high T$_b$ of C80, its sudden appearance in April 2007, and its apparent stationarity. It appears that some other intrinsic process in the jet, capable of providing a local burst in particle and/or magnetic field energy, may be responsible for the enhanced brightness temperature observed in C80. Further mid--frequency VLBA observations, currently under way, should provide the kinematical and flux evolution information necessary to obtain a better understanding of the nature of C80.

\acknowledgements This research has been supported in part by the Spanish Ministerio de Ciencia e Innovaci\'on grant AYA2007-67627-C03-03, the regional government of Andalucia grant P09-FQM-4784, and by the U.S. National Science Foundation grant AST-0907893. I.\ A. acknowledges support by an I3P contract by the Spanish Consejo Superior de Investigaciones Cient\'{\i}ficas. We thank the anonymous referee for helpful comments that improved significantly our manuscript. The VLBA is an instrument of the National Radio Astronomy Observatory, a facility of the National Science Foundation operated under cooperative agreement by Associated Universities, Inc. This research has made use of data from the MOJAVE database that is maintained by the MOJAVE team (Lister et al., 2009, AJ, 137, 3718). This research has made use of data from the University of Michigan Radio Astronomy Observatory which has been supported by the University of Michigan and by a series of grants from the National Science Foundation, most recently AST-0607523.

{\it Facilities:} \facility{VLBA ()},\facility{VLA ()}

\end{document}